\documentclass[conference]{IEEEtran}
\IEEEoverridecommandlockouts

\usepackage{cite}
\usepackage{amsmath,amssymb,amsfonts}
\usepackage{algorithmic}
\usepackage{graphicx}
\usepackage{textcomp}
\usepackage{xcolor}
\definecolor{customblue}{RGB}{0, 255, 255}  
\usepackage{hyperref}
\hypersetup{
    colorlinks=false,         
    pdfborder={0 0 1},        
    citebordercolor=green,    
    linkbordercolor=red,      
    menubordercolor=red,      
    urlbordercolor=customblue       
}

\usepackage{float}
\usepackage{textcomp}
\usepackage{booktabs} 
\usepackage{graphicx} 
\usepackage{siunitx} 
\usepackage{amsmath}
\usepackage{array} 
\def\BibTeX{{\rm B\kern-.05em{\sc i\kern-.025em b}\kern-.08em
    T\kern-.1667em\lower.7ex\hbox{E}\kern-.125emX}}
\begin{document}

\title{Semantic–Emotional Resonance Embedding: A Semi-Supervised Paradigm for Cross-Lingual Speech Emotion Recognition}

\author{Ya Zhao$^{1,2,3,4,5}$, Yinfeng Yu$^{1,2,3,4,5}$$^{, \dagger}$, Liejun Wang$^{1,2,3,4,5}$$^{, \dagger}$ \\
\thanks{$^{ \dagger}$Both Yinfeng Yu and Liejun Wang are corresponding authors. }%
\thanks{This study was funded by the National Natural Science Foundation of China (grant numbers 62463029, 62472368, and 62303259).}
\textsuperscript{1}School of Computer Science and Technology, Xinjiang University, Urumqi, China\\
\textsuperscript{2}Pengcheng Laboratory Xinjiang Network Node\\
\textsuperscript{3}Xinjiang Multimodal Intelligent Processing and Information Security Engineering Technology Research Center \\
\textsuperscript{4}Joint Research Laboratory for Embodied Intelligence, Xinjiang University\\
\textsuperscript{5}Joint International Research Laboratory of Silk Road Multilingual Cognitive Computing, Xinjiang University\\
{yuyinfeng@xju.edu.cn, wljxju@xju.edu.cn}
}

\maketitle

\begin{abstract}
Cross-lingual Speech Emotion Recognition (CLSER) aims to identify emotional states in unseen languages. However, existing methods heavily rely on the semantic synchrony of complete labels and static feature stability, hindering low-resource languages from reaching high-resource performance. To address this, we propose a semi-supervised framework based on Semantic–Emotional Resonance Embedding (SERE), a cross-lingual dynamic feature paradigm that requires neither target language labels nor translation alignment. Specifically, SERE constructs an emotion-semantic structure using a small number of labeled samples. It learns human emotional experiences through an Instantaneous Resonance Field (IRF), enabling unlabeled samples to self-organize into this structure. This achieves semi-supervised semantic guidance and structural discovery. Additionally, we design a Triple-Resonance Interaction Chain (TRIC) loss to enable the model to reinforce the interaction and embedding capabilities between labeled and unlabeled samples during emotional highlights. Extensive experiments across multiple languages demonstrate the effectiveness of our method, requiring only 5-shot labeling in the source language.
\end{abstract}


\section{Introduction}
\label{sec:intro}

Cross-lingual speech emotion recognition is dedicated to understanding emotions in multilingual interactions\cite{A1}. Ideally, machines should be able to perceive a speaker's emotional state from intonation, rhythm, and energy—much like humans do—even when unable to comprehend the semantic meaning. Cross-lingual emotional resonance is believed to stem from the mirror neuron system's~\cite{A2} mechanism of mapping others' emotional experiences onto one's own neural representations. However, current mainstream approaches heavily rely on large amounts of target-language emotion labels for supervised learning, which is difficult to achieve in low-resource language environments, thereby limiting their practical application.

\begin{figure}[!t]  
\centering  
\includegraphics[width=0.75\linewidth]{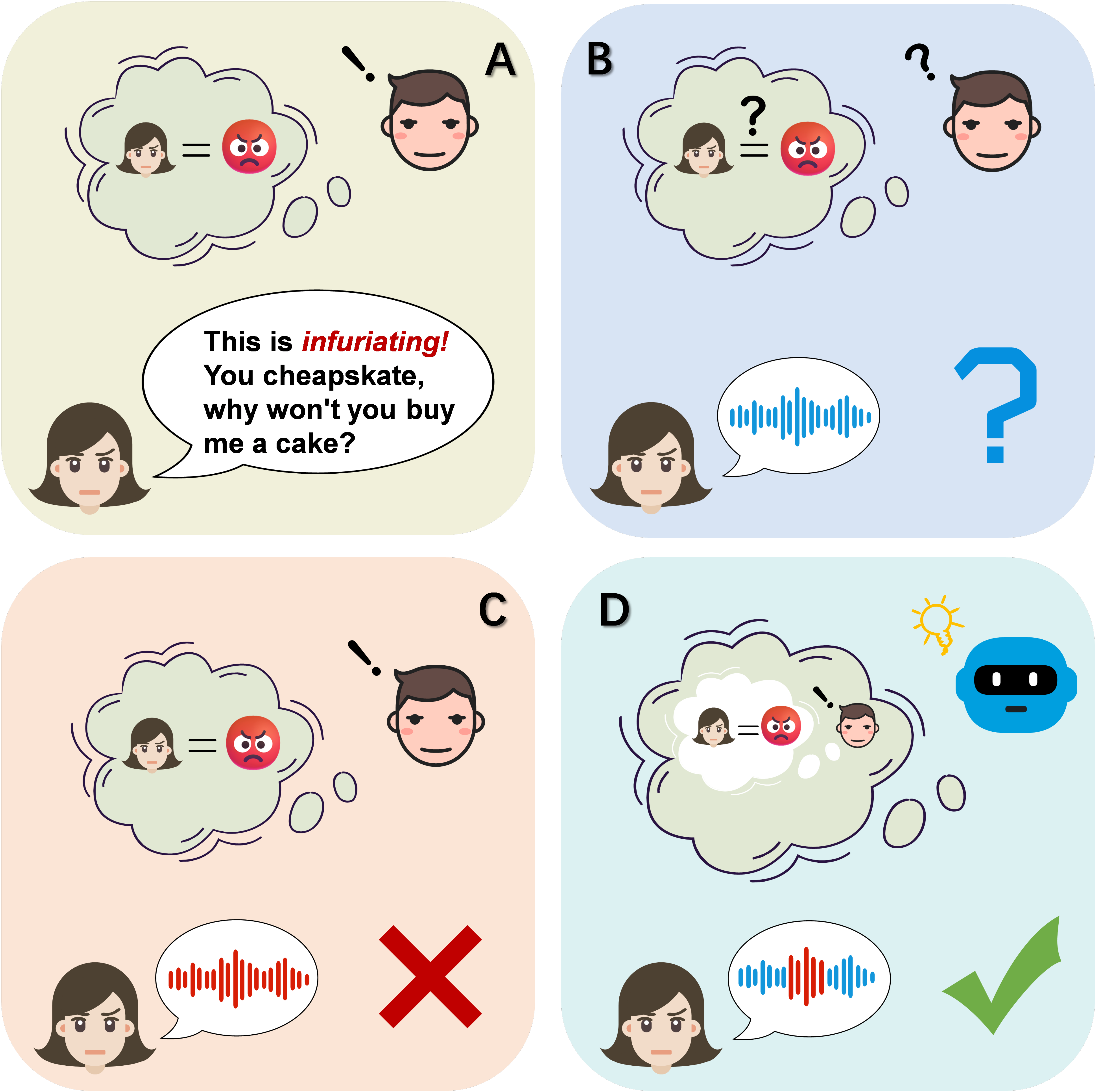}  
\caption{Traditional cross-lingual SER methods have significant drawbacks. Method A requires the presence of relevant emotional words for recognition, while B struggles with recognition due to the lack of explicit semantic content. Additionally, C relies on the assumption that the entire speech segment maintains a static emotional state. In contrast, our method D learns human emotional experiences to capture instantaneous dynamic emotional processes.}
\label{kuangjia}
\vspace{-15pt}
\end{figure}

A more fundamental challenge lies in the fact that existing CLSER methods predominantly rely on feature translation or adversarial alignment strategies\cite{A3}, implicitly assuming the existence of a language-independent emotional feature space. Such approaches typically require parallel corpora or manually aligned emotional segments, yet such data remains scarce in practice. Even when employing unsupervised alignment, cultural differences in emotional expression and acoustic diversity are often overlooked, leading to unstable cross-lingual transfer performance. In contrast, human emotional resonance requires no translation. Understanding others' emotions stems from empathy, where similar emotions across languages activate comparable neural representations. This relies on the resonance of emotional experiences rather than semantic comprehension.

Moreover, emotion is not a static attribute but a dynamic process driven by transient acoustic bursts. Existing methods rely on global statistical features or similarity metrics\cite{A55,CCC1}, making it difficult to capture the dynamic synchrony of cross-lingual speech during emotional highlights. Traditional semi-supervised learning (SSL)\cite{A5} encompasses three paradigms: pseudo-labeling\cite{A6}, consistency regularization\cite{A7}, and graph propagation\cite{A77}. All rely on translation or alignment. In contrast, CLSER's core lies in emotion-driven prosodic dynamic commonalities. These commonalities possess implicit features difficult to model through traditional alignment methods.

As shown in Fig. \ref{kuangjia}, to address the aforementioned challenges, our work proposes the SERE framework. Unlike traditional semi-supervised methods, SERE semantically anchors labeled data by learning human emotional experiences, while treating unlabeled data as a source for discovering cross-lingual affective resonance structures—rather than merely for label expansion—thus establishing a novel dynamic feature paradigm for cross-lingual semi-supervised learning. Consequently, our main contributions can be summarized as follows:
\begin{itemize}
  \item We propose an emotion resonance paradigm for cross-lingual dynamic features, leveraging language-heterogeneous encoders on unlabeled pathways to introduce a high-level semantic-combined Instantaneous Dynamic Feature Extractor (IDFE). This approach captures enhanced synchrony between transient dynamic features and emotional highlights, achieving alignment of emotions in latent space.
  \item We designed an Instantaneous Resonance Field (IRF) that focuses on emotional highlights, capturing the intensity of emotional resonance at each moment.
  \item We propose a Triple-Resonance Interaction Chain (TRIC) loss that implicitly achieves triple emotional resonance embeddings between labeled and unlabeled samples. This enhances similarity across languages during emotional experience highlights and improves the model's sensitivity to sentiment classification.
  \item Experiments conducted across 12 tasks in 4 languages demonstrate that our method achieves effective generalization with only 5-shot labeled samples.
\end{itemize}

\section{The proposed method}

This section introduces the overall architecture of our proposed SERE method.

\subsection{Overview}

\begin{figure*}[!t]  
    \centering
    \includegraphics[width=1.0\textwidth]{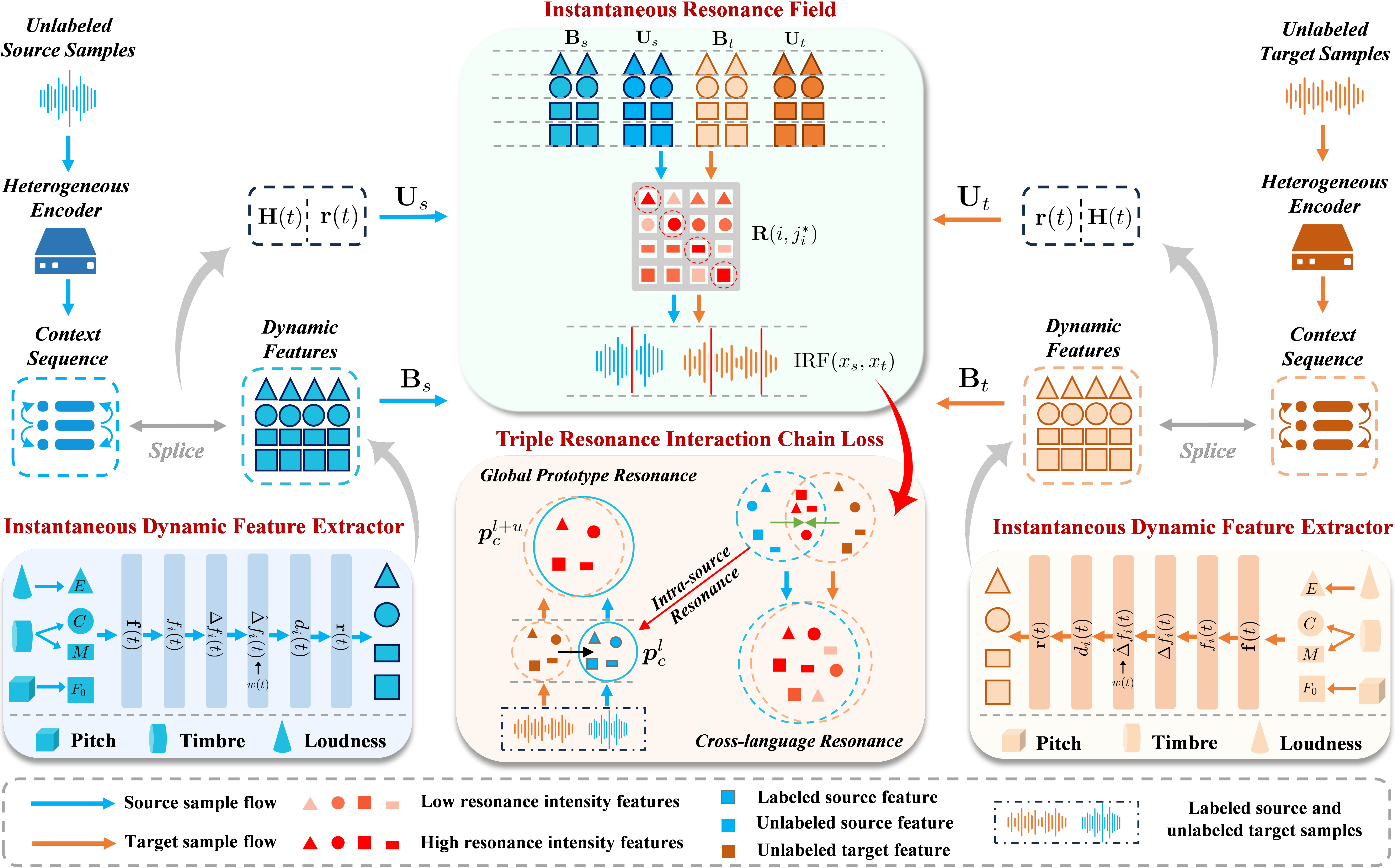}
    \vspace{-18pt}
    \caption{Overview of the proposed SERE semi-supervised dual-path architecture for CLSER tasks.}  
    \label{fig:1}  
    \vspace{-10pt}
\end{figure*}

We propose the CLSER paradigm, an efficient semi-supervised approach enabling cross-lingual resonance in emotional experiences. As shown in Fig. \ref{fig:1}, SERE incorporates two distinct data streams: a labeled path defines emotional semantic prototype anchors, while the unlabeled path pre-encodes source and target language samples using heterogeneous encoders, outputting context-aware high-level embedding sequences $\mathbf{H} = \text{E}(x) \in \mathbb{R}^{T \times d}$. Subsequently, IDFE extracts instantaneous dynamic features from the three vocal components. IRF calculates resonance similarity across linguistic emotional highlights (e.g., sudden fundamental frequency surges, abrupt energy drops), driving unsupervised self-organizing alignment among unlabeled samples. Specifically, the labeled path first defines initial prototype anchors $\boldsymbol{p}_c^l$ for emotional semantics. Unlabeled source samples automatically approach these initial anchors $\boldsymbol{p}_c^l$ based on high IRF values with labeled source samples, forming a new enhanced prototype anchor $\boldsymbol{p}_c^{l+u}$. Subsequently, this $\boldsymbol{p}_c^{l+u}$ guides unlabeled source samples toward convergence, constructing a complete emotional resonance structure within the source domain. Finally, IRF aligns target language unlabeled samples with the peak moments of source domain unlabeled samples, enabling implicit cross-lingual transfer of emotional structures. Notably, our TRIC loss designed for labeled-unlabeled interactions drives this entire emotional interaction process. This chained mechanism ensures that unlabeled data is progressively guided into embeddings, achieving semi-supervised transfer.

\subsection{Instantaneous Dynamic Feature Extractor}

In traditional methods, single static features\cite{A55} like spectral characteristics are often extracted, which easily leads to the loss of important speech features. To address this, we will utilize an Instantaneous Dynamic Feature Extractor (IDFE) to extract instantaneous dynamic features from the three fundamental elements of speech (Pitch, Loudness, and Timbre). This will result in an enhanced representation comprising four types of features. Specifically, we will extract the fundamental frequency $F_0(t)$ from the pitch. For loudness, we continue to use the RMS energy envelope $E(t)$. Regarding timbre elements, we employ the second-dimensional MFCC $M(t)$, which reflects spectral tilt and characterizes the timbre contour, along with the spectral centroid $C(t)$, which reflects the brightness of the sound. These four static features are collectively denoted as $\mathbf{f}(t) = \left[ F_0(t), E(t), M(t), C(t) \right]^\top \in \mathbb{R}^4$. Then, for each feature $f_i(t) \ (i = 1 , 2, 3, 4)$, calculate the difference between adjacent frames, where $i = 1, 2, 3, 4$ correspond to $F_0(t)$, $E(t)$,$M(t)$, $C(t)$ respectively, yielding the change amount $\Delta f_i(t) = f_i(t) - f_i(t-1)$. To enable comparison of variation across different features, we divide the difference value by its average absolute variation. First, we calculate the 
average variation amplitude of that feature across the entire speech, then normalize the current variation:
\begin{equation}
\hat{\Delta} f_i(t) = \frac{\Delta f_i(t)}{\text{Var}_i + \varepsilon},
\end{equation}
where $\text{Var}_i = \frac{1}{T} \sum_{t=1}^T \left| \Delta f_i(t) \right|$, $\epsilon = 10^{-3}$ prevents division by zero. Finally, we introduce a weight $w(t) = \sigma(\mathbf{w}^\top \mathbf{H}(t) + b)$ guided by the semantic context to amplify the dynamic signals of emotion-related frames, where $\sigma$ denotes the sigmoid function, $\mathbf{w} \in \mathbb{R}^{d}$, and $b \in \mathbb{R}$ are learnable parameters. The final dynamic feature is $d_i(t) = w(t) \cdot \hat{\Delta} f_i(t)$. By concatenating the four dynamic features as described above to obtain $\mathbf{r}(t) = \left[ d_1(t), d_2(t), d_3(t), d_4(t) \right]^\top \in \mathbb{R}^4$, and then concatenating it with the high-level semantic context sequence embedding, we form the final representation:
\begin{equation}
\mathbf{U}(t) = [\mathbf{H}(t); \mathbf{r}(t)] \in \mathbb{R}^{d+4}.
\end{equation}
This representation encompasses both diverse emotional semantics and how emotions dynamically erupt.

\subsection{Instantaneous Resonance Field}

Based on enhanced representations of the source language $\mathbf{U}_s$ and target language $\mathbf{U}_t$, we construct an Instantaneous Resonance Field (IRF) to automatically align cross-lingual emotional highlights. Specifically, we define the emotional burst intensity per frame as the weighted sum of dynamic features:
\begin{equation}
B(t) = \alpha \cdot \left| d_1(t) \right| + \beta \cdot \left| d_2(t) \right| + \gamma \cdot \left( \left| d_3(t) \right| + \left| d_4(t) \right| \right),
\end{equation}
where $\alpha, \beta, \gamma$ represents learnable intensity parameters. Next, we construct a resonance similarity matrix:
\begin{equation}
\mathbf{R}(i, j) = e^{-\delta \cdot \left(B_s(i) - B_t(j)\right)^2} \cdot \cos\left(\mathbf{U}_s(i), \mathbf{U}_t(j)\right),
\end{equation}
The first part measures cosine similarity, assessing the degree of semantic content matching, while the latter part evaluates the synchrony of burst intensity. $\delta > 0$ is the temperature coefficient, controlling burst intensity synchronization sensitivity. To preserve the emotional evolution of the source language, for each frame ${i}$ in the source language, we locate the most resonant frame $j_i^*$ in the target language—i.e., the maximum position in each row of the resonance similarity matrix. 
\begin{table*}[!t]
  \centering
  \caption{Compared with state-of-the-art CLSER experiments, where the best results are highlighted in bold.(\%)}
  \vspace{-3pt}
  \renewcommand{\arraystretch}{1.1}  
  \label{tab:method}
  \setlength{\arrayrulewidth}{0.6pt} 
  \setlength{\tabcolsep}{5.1pt}        
  \setlength{\doublerulesep}{0pt}     
    \begin{tabular}{cccccccccccccc}
    \toprule[1.2pt]
    \textbf{Method (Source$\rightarrow$Target)} & \textbf{B$\rightarrow$C} & \textbf{C$\rightarrow$B} & \textbf{B$\rightarrow$E} & \textbf{E$\rightarrow$B} & \textbf{C$\rightarrow$E} & \textbf{E$\rightarrow$C} & \textbf{B$\rightarrow$O} & \textbf{O$\rightarrow$B} & \textbf{C$\rightarrow$O} & \textbf{O$\rightarrow$C} & \textbf{E$\rightarrow$O} & \textbf{O$\rightarrow$E} & \textbf{Avg.} \\
    \hline
    JDAR \cite{B1}  & 42.70  & 48.97  & 37.95  & 47.80  & 37.58  & 35.60  & -     & -     & -     & -     & -     & -     & - \\
    JIASL \cite{B2} & 38.10  & 53.64  & 38.05  & 48.35  & 37.76  & 36.00  & -     & -     & -     & -     & -     & -     & - \\
    CIAN \cite{B3}  & 39.60  & 59.37  & 36.34  & 47.40  & 34.75  & 31.90  & -     & -     & -     & -     & -     & -     & - \\
    LRJDA-IS10 \cite{B4} & 43.10  & 50.09  & 37.57  & 48.93  & 34.88  & 38.40  & -     & -     & -     & -     & -     & -     & - \\
    ADoGT \cite{B5}  & 34.91  & 55.01  & -     & -     & -     & -     & 60.85  & 71.43  & 45.99  & 45.52  & -     & -     & - \\
    BYOL \cite{B6}  & 44.44  & 61.03  & -     & -     & -     & -     & 57.66  & 78.96  & 45.50  & 43.94  & -     & -     & - \\
    SimSiam \cite{B7} & 44.99  & 64.40  & -     & -     & -     & -     & 61.59  & 68.83  & 45.04  & 43.98  & -     & -     & - \\
    U-ERMS \cite{B8} & 47.54  & 69.04  & -     & -     & -     & -     & \textbf{61.77} & \textbf{82.35} & \textbf{50.22}  & 50.53  & -     & -     & - \\
    DAN \cite{B9}   & 36.30  & 56.72  & 33.58  & 43.50  & 32.17  & 29.30  & 35.53  & 44.24  & 36.51  & 32.42  & 32.14  & 28.93  & 36.78  \\
    NRC \cite{B10}   & 37.80  & 60.16  & 31.87  & 48.42  & 32.22  & 31.40  & 35.12  & 44.93  & 32.74  & 29.00  & 29.76  & 27.73  & 36.76  \\
    AaD \cite{B11}   & 35.00  & 55.50  & 31.47  & 48.12  & 32.17  & 29.10  & 34.52  & 44.93  & 34.92  & 27.30  & 25.60  & 26.50  & 35.43  \\
    ECAN \cite{B12}  & 39.00  & 61.37  & 34.21  & 46.87  & 34.53  & 31.90  & 36.51  & 40.86  & 35.91  & 27.42  & 28.57  & 29.16  & 37.19  \\
    \hline
    \textbf{SERE (Ours)} & \textbf{48.68} & \textbf{69.28} & \textbf{40.97} & \textbf{54.47} & \textbf{40.52} & \textbf{40.05} & 49.86  & 58.43  & 48.55  & \textbf{51.98} & \textbf{37.58} & \textbf{32.65} & \textbf{47.75} \\
    \bottomrule[1.2pt]
    \vspace{-18pt}
    \end{tabular}%
\end{table*}

We then extract enhanced representations of these optimal resonant frames from the target language and average them to obtain the source language's resonance-aware representation:
\begin{equation}
\mathbf{v}_s = \frac{1}{T_s} \sum_{i=1}^{T_s} \mathbf{U}_t \left( j_i^* \right).
\end{equation}

To obtain the overall resonance level between source and target audio, we calculate the average resonance intensity across all highlight frames. First, we extract the resonance intensity values for all highlight frames: $\mathcal{R}_{\text{max}} = \{ \mathbf{R}(i, j_i^*) \mid i = 1, 2, \ldots, T_s \}$. Then, we compute their mean as the final resonance similarity:
\begin{equation}
\text{IRF}(x_s, x_t) = \frac{1}{T_s} \sum_{i=1}^{T_s} \mathbf{R}(i, j_i^*).
\end{equation}
It is worth noting that IRF is not only used for cross-lingual emotional resonance in unlabeled data but also for emotional resonance across various types of labels.

\subsection{Triple-Resonance Interaction Chain (TRIC) Loss}

This section introduces three instances of semantic and affective resonance embedding guided by TRIC. TRIC encompasses global prototype resonance embedding, intra-source resonance embedding, and cross-lingual resonance embedding.

\paragraph{Global Prototype Resonance} This path calibrates the entire emotional semantics, where labeled source samples and unlabeled target samples will define the enhanced global prototype anchors for emotional semantics. We first compute initial prototype anchors $\boldsymbol{p}_c^l$ for each emotion category $c$ using labeled source samples $x_s^l$. For each unlabeled target sample  $x_t^u$, we calculate its IRF values with all labeled samples and select the labeled sample with the highest IRF as the pseudo-anchor $\hat{x}_s^l = \arg\max_{x_s^l \in \mathcal{D}_s^l} \text{IRF}(x_t^u, x_s^l)$. Next, we use the semantic embedding $\mathbf{z}^{\hat{x}_s^l}$ of this pseudo-anchor to represent the pseudo-semantics of the unlabeled target. Finally, we perform a weighted average of the representations of the labeled source samples and the unlabeled target samples to obtain the final enhanced global prototype anchors:
\begin{equation}
\boldsymbol{p}_c^{l+u} = \frac{1}{N_c + M_c} \left( \sum_{i:y_i=c} \mathbf{z}_i^{x_s^l} + \sum_{j:\hat{y}_j=c} \mathbf{z}_j^{\hat{x}_s^l} \right),
\end{equation}
where $N_c$ denotes the number of labeled samples in category $c$, $M_c$ represents the number of unlabeled target samples assigned to category $c$, $\hat{y}_j=c$ is the semantic embedding of the jth unlabeled sample assigned to category $c$ via IRF, $\mathbf{z}_i^{x_s^l}$ is the semantic embedding of the labeled sample, and $\mathbf{z}_j^{\hat{x}_s^l}$ is the semantic embedding of the corresponding pseudo-anchor for that sample. Building upon enhanced global prototype anchors, we introduce the global prototype resonance loss:
\begin{equation}
\begin{split}
\mathcal{L}_{\text{proto}} &= \frac{1}{C} \sum_{c=1}^{C} \left( 
    \frac{1}{N_c} \sum_{i:y_i=c} \|\mathbf{z}_i^{x_s^l} - \boldsymbol{p}_c^{l+u}\|^2 \right. \\
    &\quad + \left. \frac{1}{M_c} \sum_{j:\hat{y}_j=c} \|\mathbf{v}_j^{x_t^u} - \boldsymbol{p}_c^{l+u}\|^2 
\right),
\end{split}
\end{equation}
where $\mathbf{v}_j^{x_t^u}$ is the resonant perception representation of the jth unlabeled sample. This will compute the Euclidean distance between samples of different categories, ensuring all samples cluster closely around their corresponding enhanced prototype anchors within a unified space, achieving strong clustering.
\paragraph{Dual Instance Resonance} Building upon the enhanced global prototype anchor that achieves resonant embeddings for labeled source samples and unlabeled target samples, we define the dual instance resonance loss to realize both intra-source resonance and cross-language highlight resonance:
\begin{equation}
\mathcal{L}_{\text{dual}} = \mathbb{E}_{x^u} \left[ \left(1 - \text{IRF}(x^u, x^{\text{ref}})\right) \cdot \|\mathbf{v}^u - \mathbf{v}^{\text{ref}}\|^2 \right],
\end{equation}
where $\mathbb{E}_{x^u}$ denotes averaging over all unlabeled samples. $x^u$ represents any unlabeled sample (source or target language). $x^{\text{ref}}$ denotes its corresponding reference sample. When $x^u$ is an unlabeled source sample, $x^{\text{ref}}$ is a labeled source sample, enabling intra-source resonance. When $x^u$ is an unlabeled target sample, $x^{\text{ref}}$ is an unlabeled source sample, enabling cross-lingual highlight resonance. $\mathbf{v}^u$ and $\mathbf{v}^{\text{ref}}$ denote the corresponding resonance-aware representations.

\paragraph{Overall Objective} The overall objective of the proposed method can be defined as follows:
\begin{equation}
\mathcal{L}_{\text{SERE}} = \mathcal{L}_{\text{TRIC}} = \lambda_1\mathcal{L}_{\text{proto}} + \lambda_2 \mathcal{L}_{\text{dual}}.
\end{equation}

\section{Experiments and Results}

\subsection{Corpus and Experimental Setup}

\paragraph{Corpus} We evaluated our method across four languages (German, English, Chinese, and Italian) using publicly 
\begin{table*}[!htbp]
  \centering
  \caption{Ablation Study of the Proposed SERE Method. ‘w/o’denotes removing the component.(\%)}
  \vspace{-3pt}
  \renewcommand{\arraystretch}{1.1}  
  \label{tab:method_comparison}
  \setlength{\arrayrulewidth}{0.6pt} 
  \setlength{\tabcolsep}{5.1pt}        
  \setlength{\doublerulesep}{0pt}     
    \begin{tabular}{lcccccccccccccc}
    \toprule[1.2pt]
    \textbf{Method (Source $\rightarrow$ Target)} & \textbf{B$\rightarrow$C} & \textbf{C$\rightarrow$B} & \textbf{B$\rightarrow$E} & \textbf{E$\rightarrow$B} & \textbf{C$\rightarrow$E} & \textbf{E$\rightarrow$C} & \textbf{B$\rightarrow$O} & \textbf{O$\rightarrow$B} & \textbf{C$\rightarrow$O} & \textbf{O$\rightarrow$C} & \textbf{E$\rightarrow$O} & \textbf{O$\rightarrow$E} & \textbf{Avg.} \\
    \hline
    SERE w/o $\mathcal{L}_{\text{proto}}$ \&   $\mathcal{L}_{\text{dual}}$   & 41.77  & 65.05  & 36.00  & 51.45  & 32.14  & 32.40  & 43.12  & 51.35  & 39.54  & 41.60  & 35.66  & 30.03  & 41.68  \\
    SERE w/o $\mathcal{L}_{\text{dual}}$   & 42.36  & 65.22  & 35.78  & 49.73  & 35.60  & 35.00  & 46.98  & 55.65  & 43.37  & 43.09  & 35.74  & 31.52  & 43.34  \\
    SERE w/o $\mathcal{L}_{\text{proto}}$   & 44.90  & 67.50  & 37.67  & 51.12  & 38.17  & 37.10  & 46.52  & 55.03  & 45.29  & 46.50  & 36.80  & 32.47  & 44.92  \\
    \textbf{SERE} & \textbf{48.68} & \textbf{69.28} & \textbf{40.97} & \textbf{54.47} & \textbf{40.52} & \textbf{40.05} & \textbf{49.86}  & \textbf{58.43}  & \textbf{48.55}  & \textbf{51.98} & \textbf{37.58} & \textbf{32.65} & \textbf{47.75} \\
    \bottomrule[1.2pt]
    \vspace{-18pt}
    \end{tabular}%
\end{table*}
\begin{figure*}[!htbp]
    \centering
    \begin{minipage}[b]{0.212\textwidth}
        \centering
        \includegraphics[width=3.5cm]{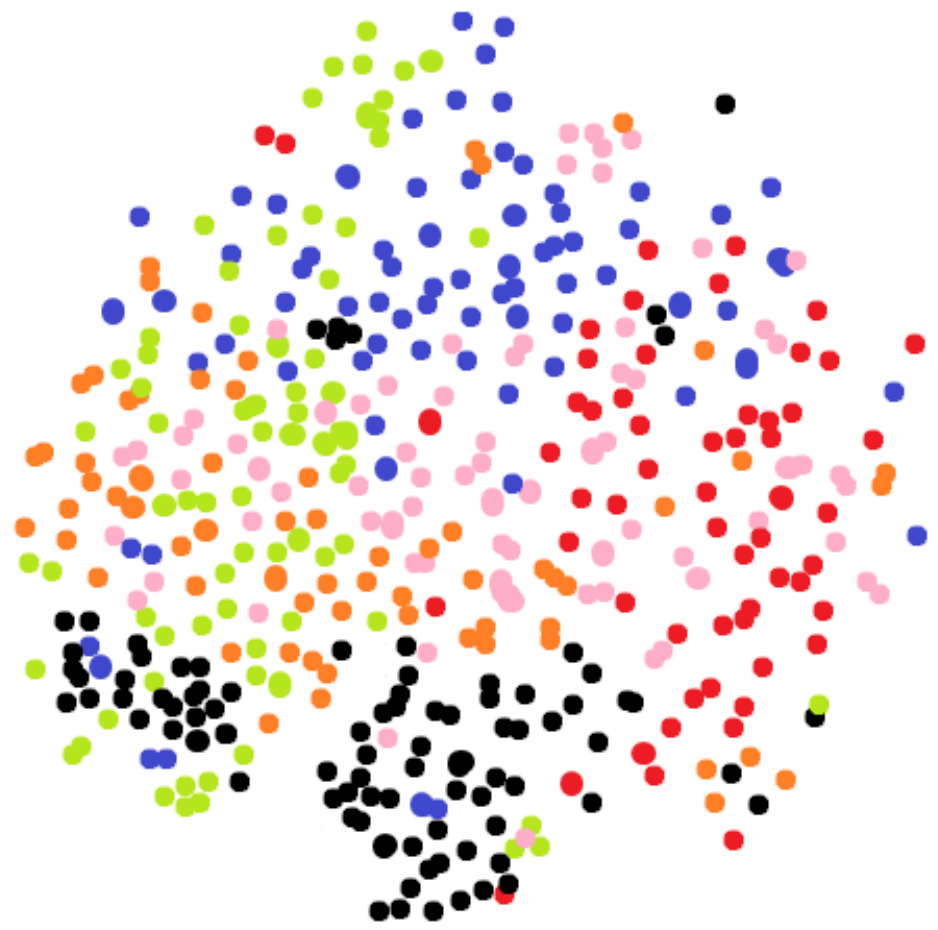}
        \vspace{2pt}
        {\footnotesize (a) SERE w/o $\mathcal{L}_{\text{proto}}$ \&   $\mathcal{L}_{\text{dual}}$}  
    \end{minipage}
    \hfill
    \begin{minipage}[b]{0.212\textwidth}
        \centering
        \includegraphics[width=3.5cm]{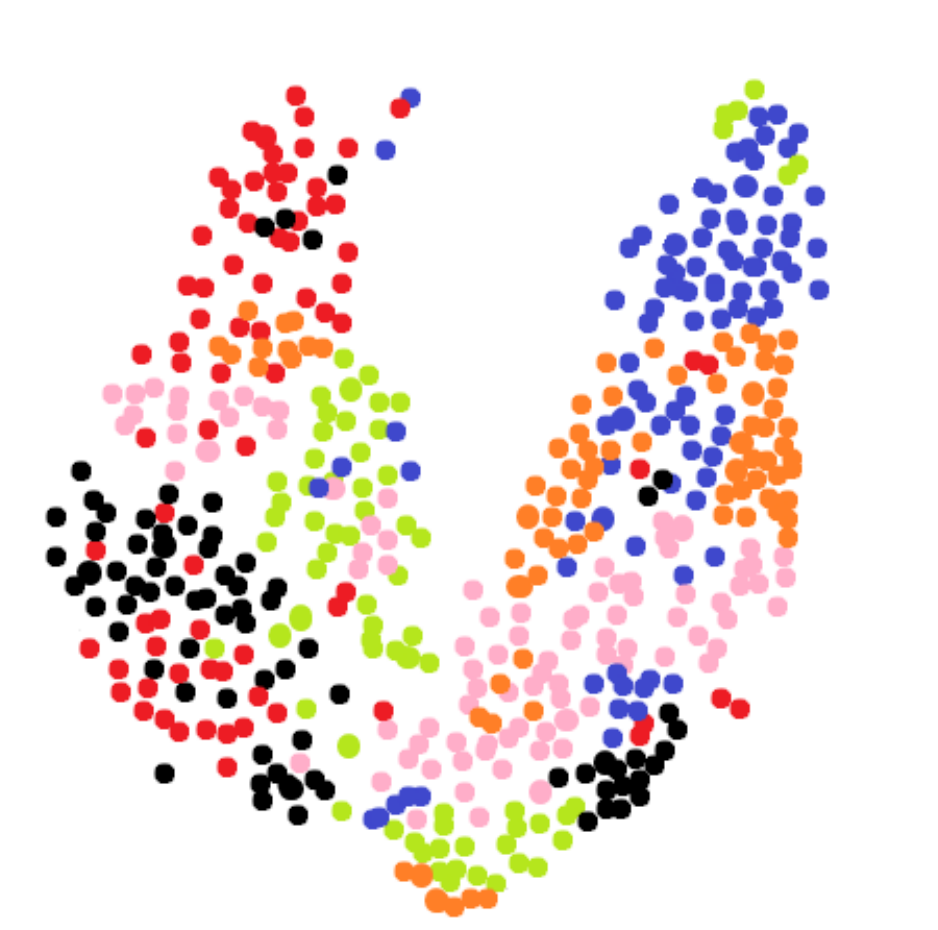}
        \vspace{2pt}
        {\footnotesize (b) SERE w/o $\mathcal{L}_{\text{dual}}$}
    \end{minipage}
    \hfill
    \begin{minipage}[b]{0.212\textwidth}
        \centering
        \includegraphics[width=3.5cm]{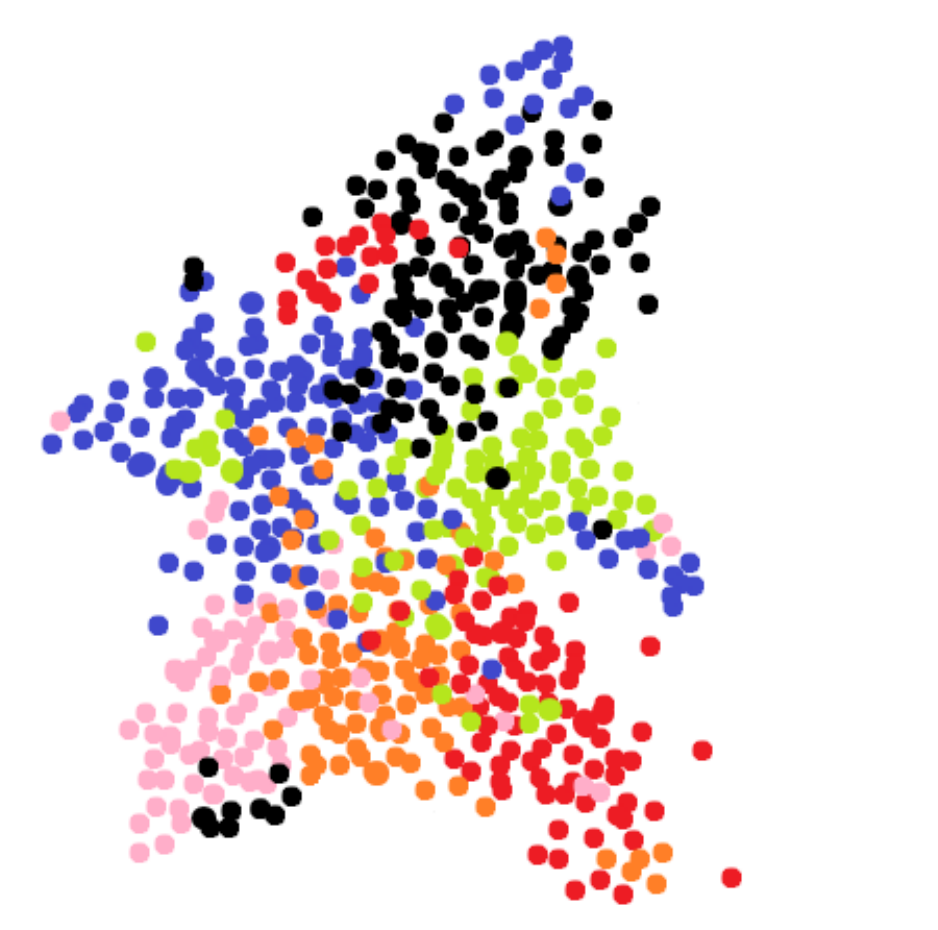}
        \vspace{2pt}
        {\footnotesize (c) SERE w/o $\mathcal{L}_{\text{proto}}$}
    \end{minipage}
    \hfill
    \begin{minipage}[b]{0.212\textwidth}
        \centering
        \includegraphics[width=3.5cm]{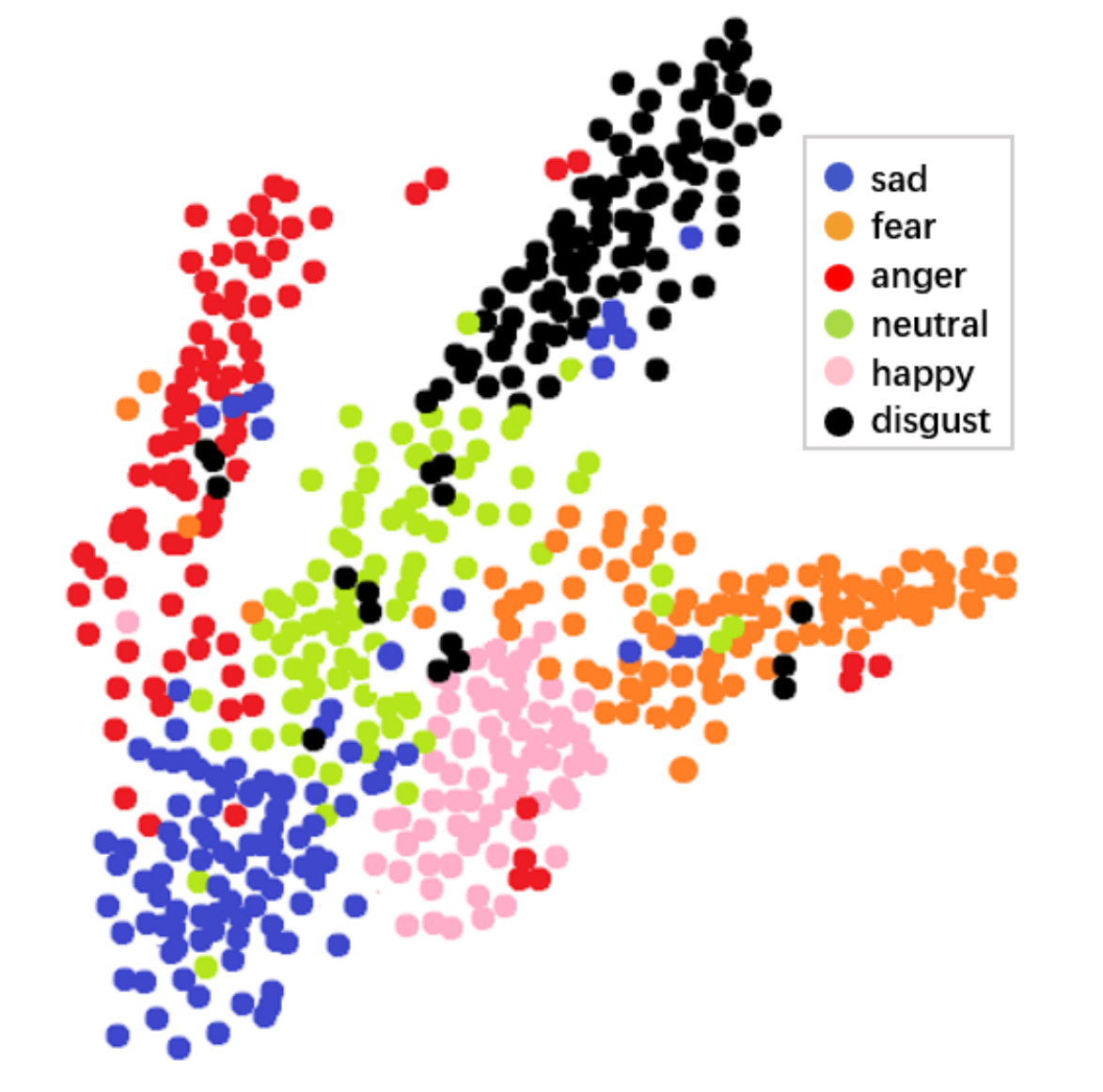}
        \vspace{2pt}
        {\footnotesize (d) SERE}
    \end{minipage}
    \vspace{-5pt}
    \small \caption{Feature distribution of different SERE components under task C$\rightarrow$B.}
    \vspace{-10pt}
    \label{fig:feature_vis}
\end{figure*}
available emotional speech corpora including EmoDB (B) \cite{C1}, eNTERFACE (E) \cite{C2}, CASIA (C) \cite{C3}, and EMOVO (O) \cite{C4}. EmoDB contains 535 German samples from 10 speakers, covering 7 emotions: neutral, anger, fear, happiness, sadness, disgust, and boredom. eNTERFACE is an English audiovisual emotion database with 1,582 samples recorded by 43 participants, covering 6 emotions: anger, disgust, fear, happiness, sadness, and surprise. CASIA is a Chinese emotional speech corpus comprising 1,200 recordings from 4 speakers, covering 6 emotions: anger, sadness, fear, happiness, neutral, and surprise. EMOVO is an Italian speech database comprising 588 samples recorded by 6 speakers, covering 7 emotions: happiness, sadness, anger, fear, disgust, surprise, and neutral.

\begin{table}[!htbp]  
  \centering
  \caption{Heterogeneous vs. Isomorphic Encoding: Performance on Unlabeled Paths.}  
  \vspace{-4pt}
  \renewcommand{\arraystretch}{1.0}  
  \setlength{\tabcolsep}{7.0pt}        
  \label{tab:encoder_comparison}    
  \setlength{\arrayrulewidth}{0.6pt} 
    \begin{tabular}{lccccc}
    \toprule[1.5pt]
    \multicolumn{2}{c}{\textbf{Method}} & \textbf{Encoders(Source$\rightarrow$Target)} & \textbf{Encoding Type} & \textbf{UA(\%)} \\
    \midrule
    \multicolumn{2}{c}{} & C-Hubert$\rightarrow$C-Hubert & Isomorphic & 46.21 \\
    \multicolumn{2}{c}{\textbf{B$\rightarrow$C}} & G-Whisper$\rightarrow$G-Whisper & Isomorphic & 46.05 \\
    \multicolumn{2}{c}{} & \textbf{   G-Whisper$\rightarrow$C-Hubert} & \textbf{Heterogeneous} & \textbf{48.68} \\
    \midrule
    \multicolumn{2}{c}{} & G-Whisper$\rightarrow$G-Whisper & Isomorphic & 66.28 \\
    \multicolumn{2}{c}{\textbf{C$\rightarrow$B}} & C-Hubert$\rightarrow$C-Hubert & Isomorphic & 66.95 \\
    \multicolumn{2}{c}{} & \textbf{C-Hubert$\rightarrow$G-Whisper} & \textbf{Heterogeneous} & \textbf{69.28} \\
    \midrule
    \multicolumn{2}{c}{} & I-Wav2vec2$\rightarrow$I-Wav2vec2 & Isomorphic & 36.77 \\
    \multicolumn{2}{c}{\textbf{E$\rightarrow$O}} & E-WavLM$\rightarrow$E-WavLM & Isomorphic & 36.65 \\
    \multicolumn{2}{c}{} & \textbf{E-WavLM$\rightarrow$I-Wav2vec2} & \textbf{Heterogeneous} & \textbf{37.58} \\
    \midrule
    \multicolumn{2}{c}{} & E-WavLM$\rightarrow$E-WavLM & Isomorphic & 31.49 \\
    \multicolumn{2}{c}{\textbf{O$\rightarrow$E}} & I-Wav2vec2$\rightarrow$I-Wav2vec2 & Isomorphic & 32.16 \\
    \multicolumn{2}{c}{} & \textbf{ I-Wav2vec2$\rightarrow$E-WavLM} & \textbf{Heterogeneous} & \textbf{32.65} \\
    \bottomrule[1.5pt]
    \vspace{-20pt}
    \end{tabular}%
\end{table}%

\paragraph{Experimental Setup} For the above corpus, we designed 12 cross-lingual SER tasks (Source corpus$\rightarrow$Target corpus). In the unlabeled path, we selected the best-performing pre-trained models for each language and performed partial fine-tuning to extract context-aware embedding sequences: For German, we utilize whisper-large-v3 (G-Whisper), while English employs wavlm-large (E-WavLM), Chinese employs hubert-base (C-Hubert), and Italian employs wav2vec2-large (I-Wav2vec2). The labeled path utilizes wav2vec2.0-base \cite{A8} to generate high-level contextual embeddings. It is particularly important to note that the output dimensions of the encoders for all paths remain consistent. Additionally, we extracted four static features from raw audio based on the three fundamental elements of speech: fundamental frequency estimated via CREPE \cite{A9}, RMS energy envelope, spectral centroid, and the 2nd-dimensional MFCC feature representing timbre contour from the Librosa toolbox \cite{A10}. These initial static features were transformed into instantaneous dynamic features through IDFE. Each training process is set to 80 epochs, using an initial learning rate of $10^{-4}$ with the Adam optimizer. To fully leverage the limited dataset, we employ 5-fold cross-validation to consistently evaluate model generalization performance. Additionally, we utilize unweighted average recall (UAR) as the evaluation metric.

\subsection{Cross-Language SER Results Comparison and Analysis}

To demonstrate the effectiveness of our proposed method, we compared it with two domain adaptation baselines, DAN \cite{B9} and AaD\cite{B11}. Table \ref{tab:method} presents the comparative results against state-of-the-art methods. SERE achieves an average UAR of 47.75\% across 12 CLSER tasks, outperforming state-of-the-art methods on 9 tasks. However, tasks including eNTERFACE–CASIA and eNTERFACE–EMOVO present significant challenges, primarily due to cultural and emotion-inducing differences between Chinese and English, as well as structural variations in vocabulary, tense, and syntax between Italian and English. Notably, our method achieves breakthroughs in the more challenging tasks C$\rightarrow$E (40.52\%), E$\rightarrow$C (40.05\%), E$\rightarrow$O (37.58\%), and O$\rightarrow$E (32.65\%), demonstrating significant advantages over existing methods and bridging emotional resonance across different languages.

Upon closer examination of the additional results, in both the B$\rightarrow$O and O$\rightarrow$B tasks, the U-ERMS\cite{B8} algorithm simultaneously reconstructs masks in both the source and target domains. This effectively prevents capturing sentiment-irrelevant information, improving performance. Regarding limitations in this task, we observed that despite linguistic similarities and belonging to the same language family, German and Italian exhibit distinct differences in emotional pronunciation. German intonation tends to be more abrupt, while Italian emotional expression is smoother and more melodic. This led our model to misclassify certain emotions. Nevertheless, our method still outperformed the average.

\subsection{Ablation Study}

We conduct ablation experiments on 12 cross-lingual tasks to validate the effectiveness of the TRIC loss. The results in Table \ref{tab:method_comparison} show that the worst performance occurs when both $\mathcal{L}_{\text{proto}}$ and $\mathcal{L}_{\text{dual}}$ are removed simultaneously, indicating that the model almost loses its cross-sample emotional resonance capability without these two components. Further comparison reveals that “w/o $\mathcal{L}_{\text{proto}}$” outperforms “w/o $\mathcal{L}_{\text{dual}}$”, suggesting that $\mathcal{L}_{\text{proto}}$ primarily enhances the embedding of single-instance emotional resonance by reducing the emotional distance between labeled source samples and unlabeled target samples, while $\mathcal{L}_{\text{dual}}$ promotes the embedding of dual-instance emotional resonance by facilitating emotional mapping both within the source domain and across domains.

To visually observe the impact of each component, we also visualized the features learned by different SERE branches for the C$\rightarrow$B task in Fig. \ref{fig:feature_vis}. The results reveal: Firstly, as shown in Fig. \ref{fig:feature_vis}(a), in the initial state without any active SERE branches, target features across emotion categories exhibit distinct, scattered distributions. Secondly, Fig. \ref{fig:feature_vis}(b) displays the outcome after removing $\mathcal{L}_{\text{dual}}$. Due to the absence of dual emotional resonance within the source language domain and across language samples, features remain dispersed, even showing fragmented distributions within the same category. Furthermore, Fig. \ref{fig:feature_vis}(c) depicts the scenario without $\mathcal{L}_{\text{proto}}$. Although target features begin to cluster, different emotion categories overlap, resulting in blurred boundaries. Finally, Fig. \ref{fig:feature_vis}(d) presents the representation under the complete SERE framework, where target features form a tighter structure with clearer category boundaries.

Table \ref{tab:encoder_comparison} validates the effectiveness of integrating language heterogeneous encoders with our framework in the unlabeled path. To more intuitively compare the performance differences between isometric encoders (where source and target languages share the same encoder) and heterogeneous encoders (where different languages use independent encoders), we conducted evaluations across four tasks. Results show that the isometric encoder significantly underperforms the heterogeneous encoder across all tasks, with the most pronounced gap observed in the C$\rightarrow$B and B$\rightarrow$C tasks. Notably, even under homogenous encoders, our SERE framework achieved stable performance across multiple pre-trained models, demonstrating its strong adaptability. This confirms that encoders with language adaptability can more effectively capture complex linguistic emotional features, while single encoders tend to conflate emotional differences across languages, further validating the advantages of language-heterogeneous encoders.

\section{Conclusion}

This paper proposes a semi-supervised CLSER paradigm called SERE that learns human emotional experiences. We introduce a heterogeneous encoder to extract high-level semantics, thereby mitigating the limitation of mismatched low-level features across languages. Under the influence of IRF, labeled and unlabeled paths achieve semantic guidance of instantaneous dynamic features and resonance with unlabeled emotional experience structures. Furthermore, our proposed TRIC loss enhances emotional clustering and linkage capabilities for low-label languages. The effectiveness of the SERE framework is validated across multiple low-resource languages with only 5-shot labeling, outperforming existing methods on several CLSER tasks.


\bibliographystyle{IEEEtran}
\bibliography{icme2026references}

@article{A1,
  title={Self supervised adversarial domain adaptation for cross-corpus and cross-language speech emotion recognition},
  author={Latif, Siddique and Rana, Rajib and Khalifa, Sara and others},
  journal={IEEE Transactions on Affective Computing},
  volume={14},
  pages={1912--1926},
  year={2022},
}

@article{A2,
  title={Is the putative mirror neuron system associated with empathy? A systematic review and meta-analysis},
  author={Bekkali, Soukayna and Youssef, George J and Donaldson, Peter H and others},
  journal={Neuropsychology review},
  volume={31},
  pages={14--57},
  year={2021},
}

@article{A3,
  title={Coarse alignment of topic and sentiment: A unified model for cross-lingual sentiment classification},
  author={Wang, Deqing and Jing, Baoyu and Lu, Chenwei and others},
  journal={IEEE Transactions on Neural Networks and Learning Systems},
  volume={32},
  number={2},
  pages={736--747},
  year={2020},
}

@article{A55,
  title={Leveraging Cross-Attention Transformer and Multi-Feature Fusion for Cross-Linguistic Speech Emotion Recognition},
  author={Zhao, Ruoyu and Jiang, Xiantao and Yu, F Richard and others},
  journal={IEEE Internet of Things Journal},
  year={2025},
}

@article{A5,
  title={Mixmatch: A holistic approach to semi-supervised learning},
  author={Berthelot, David and Carlini, Nicholas and Goodfellow, Ian and others},
  journal={Advances in neural information processing systems},
  volume={32},
  year={2019}
}

@inproceedings{A6,
  title={Pseudo-label: The simple and efficient semi-supervised learning method for deep neural networks},
  author={Lee, Dong-Hyun and others},
  booktitle={Workshop on challenges in representation learning, ICML},
  volume={3},
  number={2},
  pages={896},
  year={2013},
  organization={Atlanta}
}

@article{A7,
  title={Semi-supervised semantic segmentation with prototype-based consistency regularization},
  author={Xu, Haiming and Liu, Lingqiao and Bian, Qiuchen and others},
  journal={Advances in neural information processing systems},
  volume={35},
  pages={26007--26020},
  year={2022}
}

@article{A77,
  title={Graph random neural networks for semi-supervised learning on graphs},
  author={Feng, Wenzheng and Zhang, Jie and Dong, Yuxiao and others},
  journal={Advances in neural information processing systems},
  volume={33},
  pages={22092--22103},
  year={2020}
}

@InProceedings{A8,
  author = 	 "Baevski, Alexei and Zhou, Yuhao and Mohamed, Abdelrahman and others",
  title =        "wav2vec 2.0: A framework for self-supervised learning of speech representations",
  booktitle =        "Advances in neural information processing systems",
  year = 	 "2020",
  volume = 	 "33",
  pages = 	 "12449--12460"
}

@inproceedings{A9,
  title={Crepe: A convolutional representation for pitch estimation},
  author={Kim, Jong Wook and Salamon, Justin and Li, Peter and others},
  booktitle={2018 IEEE international conference on acoustics, speech and signal processing (ICASSP)},
  pages={161--165},
  year={2018},
  organization={IEEE}
}

@article{A10,
  title={librosa: Audio and music signal analysis in python.},
  author={McFee, Brian and Raffel, Colin and Liang, Dawen and others},
  journal={SciPy},
  volume={2015},
  pages={18--24},
  year={2015}
}

@inproceedings{B1,
  title={Cross-corpus speech emotion recognition using joint distribution adaptive regression},
  author={Zhang, Jiacheng and Jiang, Lin and Zong, Yuan and others},
  booktitle={ICASSP 2021-2021 IEEE International Conference on Acoustics, Speech and Signal Processing (ICASSP)},
  pages={3790--3794},
  year={2021},
  organization={IEEE}
}

@article{B2,
  title={Implicitly aligning joint distributions for cross-corpus speech emotion recognition},
  author={Lu, Cheng and Zong, Yuan and Tang, Chuangao and others},
  journal={Electronics},
  volume={11},
  pages={2745},
  year={2022},
}

@inproceedings{B3,
  title={Classification Inconsistency Alignment Network for Cross-corpus Speech Emotion Recognition},
  author={Zhou, Xiaoyan and Li, Jiajie and Yu, Qida and others},
  booktitle={ICASSP 2025-2025 IEEE International Conference on Acoustics, Speech and Signal Processing (ICASSP)},
  pages={1--5},
  year={2025},
  organization={IEEE}
}

@article{B4,
  title={Low-rank joint distribution adaptation for cross-corpus speech emotion recognition},
  author={Li, Sunan and Lu, Cheng and Zhao, Yan and others},
  journal={Knowledge-Based Systems},
  volume={315},
  pages={113260},
  year={2025},
}

@article{B5,
  title={Adversarial domain generalized transformer for cross-corpus speech emotion recognition},
  author={Gao, Yuan and Wang, Longbiao and Liu, Jiaxing and others},
  journal={IEEE Transactions on Affective Computing},
  volume={15},
  pages={697--708},
  year={2023},
  publisher={IEEE}
}

@article{B6,
  title={Bootstrap your own latent-a new approach to self-supervised learning},
  author={Grill, Jean-Bastien and Strub, Florian and Altch{\'e}, Florent and others},
  journal={Advances in neural information processing systems},
  volume={33},
  pages={21271--21284},
  year={2020}
}

@inproceedings{B7,
  title={Exploring simple siamese representation learning},
  author={Chen, Xinlei and He, Kaiming},
  booktitle={Proceedings of the IEEE/CVF conference on computer vision and pattern recognition},
  pages={15750--15758},
  year={2021}
}

@article{B8,
  title={An adaptation framework with unified embedding reconstruction for cross-corpus speech emotion recognition},
  author={Zhang, Ruiteng and Wei, Jianguo and Lu, Xugang and others},
  journal={Applied Soft Computing},
  volume={174},
  pages={112948},
  year={2025},
}

@inproceedings{B9,
  title={Learning transferable features with deep adaptation networks},
  author={Long, Mingsheng and Cao, Yue and Wang, Jianmin and others},
  booktitle={International conference on machine learning},
  pages={97--105},
  year={2015},
  organization={PMLR}
}

@article{B10,
  title={Exploiting the intrinsic neighborhood structure for source-free domain adaptation},
  author={Yang, Shiqi and Van de Weijer, Joost and Herranz, Luis and others},
  journal={Advances in neural information processing systems},
  pages={29393--29405},
  year={2021}
}

@article{B11,
  title={Attracting and dispersing: A simple approach for source-free domain adaptation},
  author={Yang, Shiqi and Jui, Shangling and Van De Weijer, Joost and others},
  journal={Advances in Neural Information Processing Systems},
  volume={35},
  pages={5802--5815},
  year={2022}
}

@inproceedings{B12,
  title={Emotion-aware contrastive adaptation network for source-free cross-corpus speech emotion recognition},
  author={Zhao, Yan and Wang, Jincen and Lu, Cheng and others},
  booktitle={ICASSP 2024-2024 IEEE International Conference on Acoustics, Speech and Signal Processing (ICASSP)},
  pages={11846--11850},
  year={2024},
  organization={IEEE}
}

@inproceedings{C1,
  title={A database of german emotional speech.},
  author={Burkhardt, Felix and Paeschke, Astrid and Rolfes, Miriam and others},
  booktitle={Interspeech},
  volume={5},
  pages={1517--1520},
  year={2005}
}

@inproceedings{C2,
  title={The eNTERFACE'05 audio-visual emotion database},
  author={Martin, Olivier and Kotsia, Irene and Macq, Benoit and others},
  booktitle={22nd international conference on data engineering workshops (ICDEW'06)},
  pages={8--8},
  year={2006},
  organization={IEEE}
}

@inproceedings{C3,
  title={Design of speech corpus for mandarin text to speech},
  author={Zhang, JTFLM and Jia, Huibin},
  booktitle={The blizzard challenge 2008 workshop},
  year={2008}
}

@InProceedings{C4,
  author = 	 "Costantini, Giovanni and Iaderola, Iacopo and Paoloni, Andrea and others",
  title =        "EMOVO corpus: an Italian emotional speech database",
  booktitle =        "Proceedings of the ninth international conference on language resources and evaluation (LREC'14)",
  organization = "European Language Resources Association (ELRA)",
  year = 	 "2014",
  pages = 	 "3501--3504"
}

@inproceedings{CCC1,
  title={Weavenet: End-to-end audiovisual sentiment analysis},
  author={Yu, Yinfeng and Jia, Zhenhong  and others},
  booktitle={International Conference on Cognitive Systems and Signal Processing},
  pages={3--16},
  year={2021}
}

\end{document}